\documentclass[aps,floatfix,showpacs]{revtex4}
\usepackage{graphics,epsfig}
\usepackage{graphicx}
\usepackage{psfrag}

\begin{document}
\title{Particle Motion and Perturbed Dynamical System in Warped Product Spacetimes}
\author{Pinaki Bhattacharya}
\affiliation{Gopal Nagar High School, Singur 712409, West Bengal, India \\ Department of Physics, Jadavpur University, Kolkata 700 032, India}
\author{Sarbari Guha}
\affiliation{Department of Physics, St. Xavier's College (Autonomous), Kolkata 700 016, India}

\begin{abstract}
In this paper we have used the dynamical systems analysis to study the dynamics of a five-dimensional universe in the form of a warped product spacetime with a spacelike dynamic extra dimension. We have decomposed the geodesic equations to get the motion along the extra dimension and have studied the associated dynamical system when the cross-diagonal element of the Einstein tensor vanishes, and also when it is non-vanishing. In the first case, introducing the concept of an energy function along the phase path in terms of the extra-dimensional coordinate, we have examined how the energy function depends on the warp factor. The energy function has been used as a measure of the amount of perturbation caused by a brane displacement. Geometrically the effect of brane displacement is manifested in terms of a coordinate translation along the extra dimension, thereby producing a change in the geodesic motion along the extra dimension in the region close to the brane. Then we studied the geodesic motion under a conventional metric perturbation in the form of homothetic motion and conformal motion and examined the nature of critical points for a Mashhoon-Wesson-type metric. Finally we investigated the motion for null and timelike geodesics under the condition when the cross-diagonal element of the Einstein tensor is non-vanishing and examined the effects of perturbation on the critical points of the dynamical system.
\end{abstract}

\maketitle
\section{Introduction}
\bigskip
Braneworld models with large extra dimensions are being studied for several years \cite{rs1}. Among them, the model of Randall and Sundrum \cite{rs2,rs3} with a single extra dimension, has turned out to be the most popular. The warped braneworld model was proposed by Randall and Sundrum as a possible solution to the hierarchy problem between the weak and the Planck scales. In braneworld cosmology, the Standard Model (SM) fields are assumed to remain confined to a lower dimensional hypersurface, while gravity can propagate in the bulk. In view of standard cosmology, the simplest braneworld model can be constructed by embedding the four-dimensional (4D) Friedmann-Robertson-Walker (FRW) spacetime in a five-dimensional (5D) bulk \cite{rs4}. The FRW model describes a homogenous and isotropic universe with perfect fluid matter, which remains confined to the brane under low energy conditions. The exchange of matter between the brane and the bulk is represented by a nonzero cross-diagonal ($G^{t}_{y}$) term of the Einstein tensor, which corresponds to the high energy condition. A vanishing $G^{t}_{y}$ term indicates the low energy condition.

Here we considered the RSII model in which the brane is embedded in a 5D bulk having a spacelike dynamic extra dimension. Such a model is interesting for the study of geodesic motions \cite{dahia1,dahia2,dahia3,rs5a,rs5b,rs5o,rs7,rs8,rs8i,rs10,rs11,rs11a,rs11b} and in particular the relationship between the geodesics of the higher-dimensional space and those belonging to the lower-dimensional hypersurface. While studying the classical geodesic motions of test particles of both nonzero and zero rest mass in 5D warped product spaces, we can decouple the motion along the extra dimension from the motion in the 4D hypersurfaces. This enables us to apply the method of dynamical systems (DS) in the investigation of the nature of confinement as well as the stability of the confined motions with respect to small perturbations in the hypersurfaces of such 5D spaces. Dahia et al. \cite{dahia2}, showed that a qualitative analysis of the behaviour of massive and massless particles can be made in the fifth (5th) dimension from a knowledge of the warping function. In our work, using the technique of Dahia et al. \cite{dahia2}, we have described the geodesic motion along the extra dimension by splitting the motion in the extra dimension from the motion in the 4D hypersurfaces and have studied the dynamical system under unperturbed and perturbed situations for both massive and massless particles under different types of perturbation.

The dynamical systems analysis is popular \cite{rs12a,rs12b} because very few ordinary differential equations yield explicit solutions which can be expressed in finite terms. As a result, the combination of standard functions in terms of which we can express these solutions, is inadequate for accommodating the wide range of differential equations that we encounter in practice. The qualitative study of differential equations in the DS method, is concerned with the deduction of important characteristics of the solutions of differential equations without actually solving them \cite{rs12,rs13,rs14,rs15}.

Study of cosmological perturbations has been in vogue over the years, since it is widely believed that the large scale structure of our universe developed from small linear perturbations that appeared in the early universe due to gravitational instabilities \cite{Durrer}. A lot of work has been done to study the effects of such perturbations \cite{list}. In inflationary models, perturbations are considered to be generated due to quantum fluctuations of the fields. Recently, such studies have been carried out in quantum gravity and scalar field cosmology \cite{qgravnscalar}. However, the effects of such perturbations on geodesic motions have being studied only lately \cite{geodesics}, which therefore remains an open area for investigation even today.

In our work we have considered the effect of classical linear perturbations \cite{MFB} in the form of a metric perturbation. However we have classified the metric perturbation into two types. In the first case we assumed that the background remains unperturbed, it is only the brane which is perturbed due to an interaction with the graviton. The perturbation gives rise to stresses on the brane, leading to infinitesimal brane displacements along the extra dimension, thereby affecting the motion of particles near the brane. In most of the previous works \cite{new0,new,new1,new2,new3}, brane perturbation was studied in the 5D framework. But our intention is to study the 5D configuration in a phase space. In such an analysis, the quantum fluctuations can be handled directly. Here we have assumed these fluctuations to affect the phase coordinates, and have used the perturbed coordinates as a parameter to characterize the effect on particle motion. In this case the perturbation is considered to be purely local.

In the second case we consider a conventional metric perturbation, in which the background metric consists of two parts: the unperturbed part and the perturbed part, where the perturbation is assumed to be a global phenomenon. During the study of geodesic motion under this metric perturbation, we have followed the model described by Pyne and Birkinshaw \cite{rsn2}. They made a complete generalization of the Sachs-Wolfe formalism for the timelike and spacelike components of null geodesics in a metric perturbed spacetime of arbitrary background. They explained the photon redshift and described the spatial components of the perturbed photon wavevector or lensing. However, we have used their method in a different context. We have considered a 5D Riemannian manifold and considered different types of motion, conformal and homothetic -- as the result of metric perturbation. We have also examined the stability of the system through the dynamical systems analysis.

The paper is organized as follows: Having introduced the mathematical preliminaries in Section II, we studied the geodesic motion when the cross-diagonal component of the Einstein tensor vanishes, in section III. In Section IV we have introduced the concept of an energy function along the phase path in terms of the extra-dimensional coordinate and have shown how the energy function depends on the warp factor. Then we studied the motion under coordinate translation as a result of brane displacement in Section V, and have shown that the motion along the extra dimension experiences additional force as a result of brane displacement. We also evaluated the energy function for different warping functions in this section. Subsequently we studied the dynamical system under the effect of metric perturbation in section VI. We considered two types of motion as the effect of perturbation: (i) homothetic motion and (ii) conformal motion, and derived the condition to obtain the critical points. In section VII we investigated the motion for null and timelike geodesics for the case of a non-vanishing cross-diagonal component of the Einstein tensor and examined the effects of the perturbation on the nature of critical points.

\section{Mathematical preliminaries}
The geodesic motion of a test particle in a 5D bulk is described by the equation
\begin{eqnarray*}
  \frac{d^{2}Z^{A}}{d\lambda^{2}}+^{(5)}\Gamma^{A}_{BC}\frac{dZ^{B}}{d\lambda}\frac{dZ^{C}}{d\lambda}=0,
\end{eqnarray*}
where $\lambda$ is the affine parameter along the geodesics, $Z^{A}$ are the coordinates of the bulk spacetime and $^{(5)}\Gamma^{A}_{BC}$ are the 5-dimensional Christoffel symbols of the second kind defined by $^{(5)}\Gamma^{A}_{BC}=\frac{1}{2}g^{AD}(g_{DB,C}+g_{DC,B}-g_{BC,D})$. Here we shall study the motion in a warped product space. A warped product space \cite{rs5o1,rs5o2} can be described in the following way. Let there be two Riemannian (or semi-Riemannian) manifolds $(M^{m}, h)$ and $(M^{n}, \bar{h})$ of dimensions $m$ and $n$, with metrics $h$ and $\bar{h}$ respectively. Defining a smooth function $f : M^{n} \rightarrow \Re$ (henceforth referred to as the ’warping function’), we can construct a new Riemannian (or semi-Riemannian) manifold $(M,g)$ by setting $M = M^{m} \times M^{n}$, which is defined by the metric $g = e^{2f}h \oplus \bar{h}$. In this paper we shall take $M = M^{4}\times\textrm{R}$ and identify $M^{4}$ with the (3+1)-dimensional Lorentzian manifold with signature (+ - - -).

We consider the 5D line element \cite{rs5a,rs5i} given by
\begin{equation}\label{01}
dS^2 =e^{2f(y)}\left(dt^2 - R^2(t)(dr^2 + r^2d\theta^2 + r^2sin^2(\theta)d\phi^2)\right) - T^2(t)dy^2.
\end{equation}
The function $T(t)$ is the scale factor of the extra dimension at different times in the bulk and $e^{2f(y)}$ is the extra dimensional coordinate dependent warp factor.

The non-vanishing components of the 5D Einstein tensor for the space-time under consideration are
\begin{equation}\label{02}
\bar{G}^{t}_{t}=\frac{3}{e^{2f}} \left( \frac{\dot{R}^2}{R^2}  + \frac{\dot{R}}{R}\frac{\dot{T}}{T} \right) - \frac{3}{T^2}\left( 2f^{\prime 2} + f^{\prime\prime} \right),
\end{equation}
\begin{equation}\label{03}
\bar{G}^{t}_{y}= \frac{3}{e^{2f}} \left(  \frac{\dot{T}}{T}f^{\prime} \right),
\end{equation}
\begin{equation}\label{04}
\bar{G}^{y}_{t}= -\frac{3}{T^2} \left(  \frac{\dot{T}}{T}f^{\prime} \right),
\end{equation}
\begin{equation}\label{05}
\bar{G}^{y}_{y}= \frac{3}{e^{2f}} \left( \frac{\ddot{R}}{R} + \frac{\dot{R}^2}{R^2}  \right)- \frac{6f^{\prime 2}}{T^2},
\end{equation}
and
\begin{equation}\label{06}
\bar{G}^{I}_{I}=\frac{1}{e^{2f}} \left( \frac{2\ddot{R}}{R} + \frac{\dot{R}^2}{R^2} + \frac{2\dot{R}}{R}\frac{\dot{T}}{T} + \frac{\ddot{T}}{T} \right)- \frac{3}{T^2}\left( 2f^{\prime 2}+ f^{\prime\prime}\right).
\end{equation}
Above, a `dot' represents differentiation with respect to time $t$ and a `prime, stands for differentiation with respect to the fifth coordinate $y$ and $I\equiv r,\theta,\phi$.
The 5D geodesic equations are
\begin{equation}\label{07}
\frac{d^2t}{d\lambda^2}+ 2f'\left(\frac{dt}{d\lambda}\right)\left(\frac{dy}{d\lambda}\right) + \frac{\dot{T}T}{e^{2f}}\left(\frac{dy}{d\lambda}\right)^{2}+(\dot{R}R)\left(\left(\frac{dr}{d\lambda}\right)^2-r^2 \left(\frac{d\theta}{d\lambda}\right)^2-r^2\sin^2\theta \left(\frac{d\phi}{d\lambda}\right)^2\right) = 0,
\end{equation}
\begin{equation}\label{08}
\frac{d^2r}{d\lambda^2}+ 2\left(\frac{\dot{R}}{R}\right)\left(\frac{dt}{d\lambda}\right)\left(\frac{dr}{d\lambda}\right) - {2}f'\left(\frac{dr}{d\lambda}\right)\left(\frac{dy}{d\lambda}\right)+ r\left(\frac{d\theta}{d\lambda}\right)^2+ r \sin^2\theta\left(\frac{d\phi}{d\lambda}\right)^2 = 0,
\end{equation}
\begin{equation}\label{09}
\frac{d^2\theta}{d\lambda^2}+{2}\left(\frac{\dot{R}}{R}\right)\left(\frac{dt}{d\lambda}\right) \left(\frac{d\theta}{d\lambda}\right)+{2}f'\left(\frac{d\theta}{d\lambda}\right) \left(\frac{dy}{d\lambda}\right)+\frac{{2}}{r}\left(\frac{d\theta}{d\lambda}\right)\left(\frac{dr}{d\lambda}\right) -\cos\theta\sin\theta\left(\frac{d\phi}{d\lambda}\right)^2 = 0,
\end{equation}
\begin{equation}\label{10}
\frac{d^2\phi}{d\lambda^2}+{2}\left(\frac{\dot{R}}{R}\right)\left(\frac{dt}{d\lambda}\right) \left(\frac{d\phi}{d\lambda}\right)+{2}f'\left(\frac{d\phi}{d\lambda}\right)\left(\frac{dy}{d\lambda}\right) +\frac{{2}}{r}\left(\frac{d\phi}{d\lambda}\right)\left(\frac{dr}{d\lambda}\right) +{2}\cot\theta\left(\frac{d\phi}{d\lambda}\right)\left(\frac{d\theta}{d\lambda}\right) = 0,
\end{equation}
\begin{equation}\label{11}
\frac{d^2y}{d\lambda^2}+\frac{{2}\dot{T}}{T}\left(\frac{dy}{d\lambda}\right)\left(\frac{dt}{d\lambda}\right) +\frac{f'e^{2f}}{T^2}\left(\left(\frac{dt}{d\lambda}\right)^2-R^{2}\left(\left(\frac{dr}{d\lambda}\right)^2+r^2 \left(\frac{d\theta}{d\lambda}\right)^2+r^2\sin^2\theta \left(\frac{d\phi}{d\lambda}\right)^2\right)\right)= 0.
\end{equation}

Following the method adopted by Ghosh and Kar in \cite{rs11a}, here in this study we shall assume
\begin{equation}\label{21}
  \frac{\dot{R}}{R}= a,
\end{equation}
and
\begin{equation}\label{22}
  \frac{\dot{T}}{T}= -c.
\end{equation}

\section{Geodesic motion when $G^{t}_{y} = 0$}
From the above we find that the cross diagonal element of the Einstein tensor (i.e. $G^{t}_{y}$) will vanish if
\begin{equation}\label{23}
 \dot{T}f^{\prime} = 0.
\end{equation}
This means that either $\dot{T}=0$, or $f'=0$. But the condition $f' = 0$ is inadmissible, since that will represent an unwarped spacetime. Therefore the scale factor of the extra dimensional coordinate must be constant, which we can assume to be $T(t)=1$. This corresponds to a static extra dimension, for which we have a stabilized bulk.

To study the motion in the 5D bulk we consider the first integral of the geodesics:
\begin{equation}\label{24}
g_{AB}\frac{dZ^A}{d\lambda}\frac{dZ^B}{d\lambda} = \varepsilon_{5},
\end{equation}
where $\varepsilon_{5}=1$ for timelike geodesics and $\varepsilon_{5}=0$ for null geodesics. Here we will concentrate on timelike and null geodesic motion along the 5th dimension and will study the nature of the critical points.
A qualitative analysis of the motion in the 5th dimension can be done without actually solving the above equations by defining $q=\frac{dy}{d\lambda}$ and investigating the corresponding dynamical system \cite{rs12a}:
\begin{equation}
\frac{dy}{d\lambda}=q \label{24a},
\end{equation}
\begin{equation}
\frac{dq}{d\lambda}=Q(q,y) \label{24b}.
\end{equation}
The \emph{equilibrium points} of the system of equations (\ref{24a}) and (\ref{24b}) are given by $\frac{dy}{d\lambda}=0$ and $\frac{dq}{d\lambda}=0$. Knowledge of these points along with their stability properties provides a lot of information about the behavior of this dynamical system.

With the help of equations (\ref{11}) and (\ref{24}) we can write
\begin{equation}\label{25}
\ddot{y}= -f'(y)(\varepsilon_{5}+\dot{y}^{2}).
\end{equation}
Here `dot' denotes differentiation with respect to the affine parameter $\lambda$ along the geodesics. We can redefine the above dynamical system in the form
\begin{equation}\label{26}
\dot{y}= q = P_{1}(q,y),
\end{equation}
and
\begin{equation}\label{27}
\dot{q}= -f'(y)(\varepsilon_{5}+q^{2})= Q_{1}(q,y).
\end{equation}
For timelike geodesics, this system may have a critical point at $q = 0$ and $y = l$, if $f'(l)= 0$. On the other hand for null geodesics, the system has an infinite number of critical points along $y$. The character of such a system was thoroughly investigated in \cite{dahia2}. To determine the stability and the eigenvalue of the system, we calculate the partial derivatives of $P_1$ and $Q_1$ with respect to $y$ and $q$ at the critical point. Thus we obtain
\begin{equation}\label{28}
   (\partial P_{1} /\partial y)|_{q= 0,y=l}= 0,
   \end{equation}
   \begin{equation}\label{29}
   (\partial P _{1}/\partial q)|_{q= 0,y=l}= 1,
   \end{equation}
   \begin{equation}\label{30}
   (\partial Q_{1} /\partial y)|_{q= 0,y=l}= -\varepsilon_5 f''(l),
   \end{equation}
   \begin{equation}\label{31}
   (\partial Q_{1} /\partial q)|_{q= 0,y=l}= 0.
\end{equation}
It can be said from the above that for null geodesics the system has infinite number of critical points along the extra dimension. On the other hand, for timelike geodesics, the nature of the critical point depends on the value of $f''(l)$.

\section{The energy function along the phase path}
Let us go back to the equations of the 5D geodesics given by (\ref{24}), which with the help of (\ref{11}) led us to (\ref{27}). As mentioned above, for timelike geodesics the system may have a critical point at $q=0$ and $y=l$ (i.e at $y=l$, $f'(l)=0$). If so, then this critical point will be a saddle point for $f''(l)<0$, and will be a center for $f''(l)>0$. On the other hand for null geodesics, the equilibrium point will lie on the line $q=0$. Whatever be the situation, we can determine the phase path for the system, which is given by the relation
\begin{equation}\label{14}
    f(y)=-\ln\frac{\sqrt{\varepsilon_{5}+q^{2}}}{K_{0}},
\end{equation}
where $K_{0}$ is the integration constant.
After some simple calculations this equation can be recast in the form
\begin{equation}\label{15}
   K e^{-2f(y)}=(\varepsilon_{5}+q^{2}),
\end{equation}
where K is a positive constant. Thus along the phase path we can write equation (\ref{25}) as
\begin{eqnarray*}
\ddot{y}=\frac{d}{dy}\left(\frac{Ke^{-2f(y)}}{2}\right).
\end{eqnarray*}

Drawing analogy with Newtonian mechanics, the quantity $\left. \frac{Ke^{-2f(y)}}{2} \right.$ can be interpreted as the potential along the phase path. Consequently we can introduce the concept of an energy function along the phase path, which can be written as
\begin{equation}\label{15a}
 E=\frac{\dot{y}^{2}}{2}+\frac{K e^{-2f(y)}}{2}.
\end{equation}
The above equation is analogous to the energy expression of a harmonic oscillator (which is often considered as a toy model to describe brane-bulk interactions \cite{BBC}). The difference is that the energy function defined above is not necessarily conserved, unlike the case of a simple harmonic oscillator. The introduction of the concept of energy function is useful because in the case of a coordinate translation due to a brane displacement, we can make explicit calculations for this energy function and estimate the change in its value, which can therefore be used as a measure of the amount of perturbation generated by the brane displacement.

Eliminating $\dot{y}$ (i.e. $q$) between (\ref{15}) and (\ref{15a}), we obtain
\begin{equation}\label{15b}
e^{2f(y)}=\frac{2K}{2E+\varepsilon_{5}}.
\end{equation}
This indicates how the energy function defined by us is related to the warp factor $e^{f(y)}$.

\section{Geodesic motion under coordinate translation}
The equation
\begin{eqnarray*}\
    \ddot{y_{0}}= -f'(y_{0})(\varepsilon_{5}+\dot{y_{0}}^{2}),
\end{eqnarray*}
represents the geodesic motion for the unperturbed system, where $y_0$ represents the location of the brane along the extra dimension in the unperturbed condition.  Let us  consider a stress on the brane arising due to the bending of the brane on account of its interaction with the graviton. This interaction gives rise to an infinitesimal translation of the brane along the extra dimension, which is now relocated at the position
\begin{equation}\label{15c}
y=y_{0}\pm \epsilon y_{1},
\end{equation}
where we have used the idea of brane perturbation described in \cite{new}. Under such a translation of the brane, the equation of geodesic motion along the 5th dimension can be written as
\begin{equation}\label{16}
    \ddot{y}= -f'(y_{0}+\epsilon y_{1})(\varepsilon_{5}+\dot{y}^{2}),
\end{equation}
where $\epsilon$ is a very small quantity and we assume $y_{1}$ to be independent of the affine parameter, absorbing the $\pm$ sign within $y_{1}$. That means we consider $y_1$ to be positive if the brane is displaced towards the positive $y$-direction and negative for the displacement along the negative $y$-direction. Using Taylor's formula to expand (\ref{16}) about the coordinate of the unperturbed brane, we have
\begin{equation}\label{17}
    \ddot{y}= -\left(f'(y_{0})+(\epsilon y_{1})f''(y_{0})+(\epsilon y_{1})^{2}f'''(y_{0})/2+......\right)(\varepsilon_{5}+\dot{y}^{2}).
\end{equation}
We find that additional terms appear on the righthand side of equation (\ref{17}) as a result of this expansion. If we restrict to first order perturbations in (\ref{17}), the additional term can be treated as an additional force term. We note that even if we put $y_{0}= 0$, still the additional force will not vanish provided $f''(y_{0})\neq 0$. The condition $f''(y_{0})= 0$ implies that the critical point is degenerate. This may happen if we choose the warping function to be a constant one, which is inadmissible in the present model.

If we consider a growing or decaying warping function i.e $f(y)= \pm a_1 \ln\cosh (y)$, then for timelike geodesics, considering the first order perturbation, equation (\ref{17}) can be written as
\begin{equation}\label{18}
 \ddot{y}= \mp a_1 \tanh(y_{0})(1+\dot{y^{2}})\mp 4a_1 \epsilon y_{1}\cosh^{-2}(y_{0})(1+\dot{y}^{2}).
\end{equation}

Let us now consider the change in the energy function along the phase path on account of coordinate translation of the brane, which can now be expressed in the form
\begin{eqnarray*}
E=E_{0}+\epsilon E_{1},
\end{eqnarray*}
and proceed to find the exact form of equation (\ref{15b}) under this perturbed condition for the different warp factors, which will tell us how the the energy function changes on account of perturbation for the different warping functions.

\subsubsection{Growing warping function}
\begin{figure}
\includegraphics[height= 1.5in]{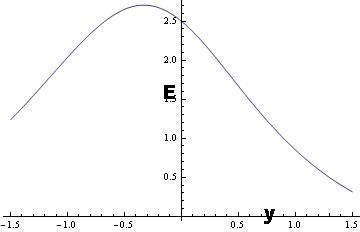}
  \caption{Variation of Energy function with extra dimension along the phase path for growing warping function under perturbed condition}
  \label{fig1}
\end{figure}

Let us first consider the warping function $f(y)=A\ln\cosh(By)$. For simplicity of calculations, we will consider $A= 1/2$ and $B=1$. Thus the equation (\ref{15b}) will look like
\begin{eqnarray*}
E=\frac{2K(\frac{1}{\cosh y})-1}{2}.
\end{eqnarray*}
Since we are considering only the 1st order perturbation, the perturbed equation for the energy function can be written as
\begin{equation}\label{19}
E_{0}+\epsilon E_{1}=\frac{2K(\frac{1}{\cosh y_{0}}-\epsilon y_{1}\frac{\sinh y_{0}}{(\cosh y_{0})^{2}} )-1}{2}.
\end{equation}
Thus
\begin{eqnarray*}
E_{0}=\frac{2K(\frac{1}{\cosh y_{0}})-1}{2}
\end{eqnarray*}
and
\begin{eqnarray*}
E_{1}=-2K y_{1}\left(\frac{\sinh y_{0}}{\cosh y_{0}^{2}}\right),
\end{eqnarray*}
where $E_{0}$, $E_{1}$ are the zeroth and the 1st order terms for the energy function under the perturbed condition. Fig.~\ref{fig1} demonstrates how the perturbed energy function varies along the extra dimension.

\subsubsection{Decaying warping function}
\begin{figure}
\includegraphics[height= 1.5in]{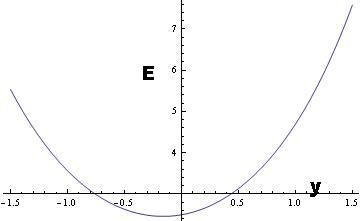}
  \caption{Variation of Energy function with extra dimension along the phase path for decaying warping function under perturbed condition}
  \label{fig2}
\end{figure}

Here we consider $f(y)=-A\ln\cosh(By)$, where A and B are positive constants. As in the previous case we choose $A=1/2$ and $B=1$. Equation (\ref{19}) now looks like
\begin{eqnarray*}
E=\frac{2K\cosh y-1}{2}.
\end{eqnarray*}
Considering only the 1st order perturbation, the perturbed equation is now
\begin{equation}\label{20}
E_{0}+\epsilon E_{1}=\frac{2K(\cosh y_{0}+\epsilon y_{1}\sinh y_{0})-1}{2}.
\end{equation}
Therefore \begin{eqnarray*}
E_{0}= \frac{2K\cosh y_{0}-1}{2},
\end{eqnarray*}
and
\begin{eqnarray*}
E_{1}= K y_{1}\sinh y_{0}.
\end{eqnarray*}
Here $E_{0}$ and $E_{1}$ have the same interpretation as in the previous case, and Fig.~\ref{fig2} demonstrates the variation of the perturbed energy function along the extra dimension.

\section{Geodesic Motion under conventional metric perturbation}


Following the usual formalism of metric perturbation theory as considered by Pyne and Birkinshaw in \cite{rsn2}, let us assume the full metric to be represented by the expression
\begin{equation}\label{33}
  g_{AB} = g^{0}_{AB}+ h_{AB},
\end{equation}
where $g^{0}_{AB}$ represents the unperturbed metric and $h_{AB}$ is the perturbed part. The basic mathematical formulation given below is the taken from \cite{rsn2}, but we have applied it to study the effect of perturbation when the nature of perturbation resembles a homothetic motion or a conformal motion.

The Christoffel symbols of the metric $g_{AB}$ given above can be split into the zeroth and first order components in $h$ as follows:
\begin{equation}\label{33a}
\Gamma^{C}_{AB}= \Gamma^{(0)C}_{AB}+\Gamma^{(1)C}_{AB},
\end{equation}
where
\begin{center}
$\Gamma^{(0)C}_{AB} = \frac{1}{2}g^{(0)CD}(g^{(0)}_{AD,B}+g^{(0)}_{BD,A}- g^{(0)}_{AB,D})$,
\end{center}
and
\begin{center}
$\Gamma^{(1)C}_{AB} = \frac{1}{2}g^{(0)CD}(h_{AD;B}+h_{BD;A}- h_{AB;D})$.
\end{center}
As in \cite{rsn2}, we assume that $x^{(0)C}(\lambda)$ represents the geodesic of the bulk spacetime with $\lambda$ as the affine parameter. We will refer to $x^{(0)C}(\lambda)$ as the ``unperturbed path'', which satisfies the geodesic equation in the unperturbed bulk. The equation for the unperturbed geodesic is then given by
\begin{equation}\label{34}
\ddot{x}^{(0)C}+\Gamma^{(0)C}_{AB}(x^{0})\dot{x}^{(0)A}\dot{x}^{(0)B}= 0,
\end{equation}
where `dot' represents differentiation with respect to the affine parameter $\lambda$. We can write
\begin{equation}\label{35}
x^{C}(\lambda)= x^{(0)C}(\lambda)+x^{(1)C}(\lambda),
\end{equation}
where $x^{(1)C}(\lambda)$ describes the perturbed geodesic path.

From equations (\ref{35}) and (\ref{34}) we get
$$ \ddot x^{C}=-\Gamma^{(0)C}_{AB}\left(x^{(0)} \right) {\dot x^{(0)A}}{\dot x^{(0)B}}+\ddot x^{(1)C}. \label{(35a)}$$
To ensure that $x^{C}(\lambda )$ is an affinely parameterized geodesic of the perturbed spacetime, the following condition must be satisfied
\begin{equation}\label{35b}
\ddot x^{C}	=-\Gamma^{C}_{AB} \left( x\right) {\dot x^{A}}{\dot x^{B}} =-\Gamma^{(0)C}_{AB} \left( x\right)
\left( {\dot x^{(0)A}}{\dot x^{(0)B}}+2{\dot x^{(0)A}}{\dot x^{(1)B}}\right)	
	 -\Gamma^{(1)C}_{AB}\left( x\right){\dot x^{(0)A}}{\dot x^{(0)B}}, 	
\end{equation}
which is obtained with the help of equations (\ref{33a}) and (\ref{35}), keeping only the first order terms. If we assume that there are no singularities in the neighborhood of these geodesic paths, we can expand the connection coefficients near the perturbed path, $x$, about the unperturbed path, $x^{(0)}$, as follows:
\begin{equation}\label{35c}
 \Gamma^{(0)C}_{AB}\left( x\right)=\Gamma^{(0)C}_{AB}\left( x^{(0)} \right)+\Gamma^{(0)C}_{AB ,D} \left( x^{(0)} \right)x^{(1)D}+ \dots	\Gamma^{(1)C}_{AB}\left( x\right)= \Gamma^{(1)C}_{AB}\left( x^{(0)}\right)+\dots .
\end{equation}

\noindent
Substituting (\ref{35c}) into (\ref{35b}), we obtain
\begin{equation}\label{36}
\ddot{x^{C}}=-\Gamma^{(0)C}_{AB}(x^{0})\dot{x}^{(0)A}\dot{x}^{(0)B}-\Gamma^{(1)C}_{AB}(x^{0})\dot{x}^{(0)A}\dot{x}^{(0)B} -2\Gamma^{(0)C}_{AB}(x^{0})\dot{x}^{(0)A}\dot{x}^{(1)B}-\Gamma^{(0)C}_{AB,D}(x^{0})\dot{x}^{(0)A}\dot{x}^{(0)B}x^{(1)D}.
\end{equation}

We will now use the above formulation to study the geodesic motion along the extra dimension under the condition $G^t_y=0$. So the second term of equation (\ref{36}) will vanish and the term $x^{(1)D}$ can be written as $y^{(1)}$. Thus for the motion along the extra dimension, equation (\ref{36}) can be written as
\begin{equation}\label{37}
\ddot{y}=-\Gamma^{(0)y}_{AB}(x^{0})\dot{x}^{(0)A}\dot{x}^{(0)B}-\Gamma^{(1)y}_{AB}(x^{0})\dot{x}^{(0)A}\dot{x}^{(0)B}
-\Gamma^{(0)y}_{AB,y}(x^{0})\dot{x}^{(0)A}\dot{x}^{(0)B}y^{(1)}.
\end{equation}
We point out that the quantity $y^{(1)}$ in (\ref{37}) is different from the parameter $y_1$ introduced in (\ref{15c}).

\subsection{Perturbation as homothetic motion}
During a point transformation from one point of a manifold to another point of the manifold if the metric obeys the relation $\overline{g}= \rho^{2}g$, then the transformation will describe a homothetic motion (see \cite{rsn3}), provided $\rho$ is a constant. If the transformation takes place in a particular direction, say $\xi$, then we can write
\begin{center}
   $ L_{\xi}g_{AB}= 2\phi g_{AB}$,
\end{center}
where $\phi$ is another constant. If $\phi\neq0$, then $\phi$ is called a homothetic vector field (HVF). In this paper we have assumed that the brane is embedded in the 5D bulk and following the description given by Nash \cite{Nash}, we assume that by constantly perturbing the brane along its normal direction, we obtain a submanifold of the same bulk, provided the embedding function remains regular. So we can treat this submanifold as a regular submanifold. Let us consider that a motion has taken place between a point of the parent manifold and the corresponding submanifold. Due to the perturbation of the bulk, the metric will assume the form $g'_{AB}= g_{AB}+L_{\xi}g_{AB}$. So assuming $\xi$ to be a HVF, we can write
\begin{center}
    $g_{AB}= g^{0}_{AB}+2\phi g^{0}_{AB}$, \qquad   i.e. \qquad  $h_{AB} = 2\phi g^{0}_{AB}$.
\end{center}
Due to the homothetic motion, the first term of equation (\ref{37}) will represent the unperturbed geodesic, the second term will vanish and after some calculations, the third term can be expressed in terms of the warping function. Thus we can rewrite equation (\ref{37}) as
\begin{equation}\label{38}
\ddot{y}=-f'(y)(\varepsilon_5+\dot{y}^{2})- f''(y)(\varepsilon_5+\dot{y}^{2})y^{1}- 2f'(y)^{2}(\varepsilon_5+\dot{y}^{2})y^{1},
\end{equation}
where we have dropped the round bracket in the superscript of $y^1$. We now define the dynamical system
\begin{equation}\label{39}
\dot{y}= q,
\end{equation}
and
\begin{equation}\label{40}
\dot{q}= -f'(y)(\varepsilon_5+q^{2})-(f''(y)+2f'(y)^{2})(\varepsilon_5+q^{2})y^{1}.
\end{equation}

Assuming that $y^1$ is given by a small change $\delta y$, we can rewrite the dynamical system under this condition in the form
\begin{equation}\label{41}
\dot{y}= q,
\end{equation}
and
\begin{equation}\label{42}
\dot{q}= [-f'(y)(\varepsilon_5+q^{2})]_{unperturbed}-[(f''(y)+2f'(y)^{2})(\varepsilon_5+q^{2})\delta y]_{perturbed}.
\end{equation}
Alternatively we can write
\begin{equation}\label{43}
\dot{y}= q = P_{2}(q,y),
\end{equation}
and
\begin{equation}\label{44}
\dot{q}= [-f'(y)(\varepsilon_5+q^{2})]_{unperturbed}-[\delta(f'(y)e^{2f(y)})e^{-2f(y)}(\varepsilon_5+q^{2})]_{perturbed}= Q_{2}(y,q).
\end{equation}
From now onwards, we will denote all small changes by the conventional symbol ``$\delta$''. The first term in (\ref{44}) describes the background term i.e. the unperturbed term, whereas the second part describes the additional term due to perturbation.

The critical point of the perturbed system for timelike geodesics can be determined from the equations
\begin{eqnarray}
   P_{2}(q,y) &=& 0, \nonumber \\
   \\
    Q_{2}(q,y) &=& 0. \nonumber
    \end{eqnarray}
The first equation tells us that $q = 0$, and the second equation indicates that to get a critical point there must be a point $y = l$, such that
\begin{eqnarray}
 -f'(l)-\delta(f'(l)e^{2f(l)})e^{-2f(l)}& = &0,
\end{eqnarray}
or
\begin{eqnarray}
 -\delta(f'(l)e^{2f(l)})& = &f'(l)e^{2f(l)}.
\end{eqnarray}
This indicates that the deviation of the function $f'(y)e^{2f(y)}$ is numerically equal to the value of this function at $y=l$. But such a situation is not acceptable when the perturbation is very small. On the other hand, if we assume $f'(y)e^{2f(y)}$ to be a conservative term, then the term $\delta(f'(y)e^{2f(y)})$ will vanish, and as a result the perturbed part will also vanish. So the geodesic equation will then resemble equation (\ref{27}). In that condition, the expression for the warping function can be obtained from the condition
\begin{eqnarray}
  f'(y)e^{2f(y)} &=& \textrm{constant}.
\end{eqnarray}
A simple solution of such an equation will be given by $f(y)= \frac{1}{2}\log(Cy)$, where $C$ is a constant. This warping function will describe a 5D Riemannian space endowed with a Mashhoon-Wesson-type metric \cite{weasson}. The nature of the warping function for this metric suggests that there is no point along the extra dimension for which $f(l)=0$. In their paper, Dahia et al \cite{DGR} demonstrated that for this type of metric, there is no confinement of particles in the $y=\textrm{constant}$ hypersurfaces.

\subsection{Perturbation as conformal motion}

If a point transformation does not change the angle between two directions at a point, then the transformation is known as a conformal motion \cite{rsn3}. The symmetry condition for conformal killing vectors can be written as
\begin{center}
   $ L_{\xi}g_{AB}= 2\phi g_{AB}$.
\end{center}
It is known that if we choose $\phi = 0$ then $\xi$ becomes a Killing vector. If we choose $\phi = \textrm{constant} $ then it represents a homothetic vector, which we have considered in the previous subsection. But in the case of conformal motion, $\phi$ can be a function of both space coordinates and time \cite{far}. Here we assume that $\phi$ is a function of the extra-dimensional coordinate only. Under such a condition, the perturbed dynamical system for $G^t_y = 0$ can be written as
 \begin{equation}\label{45}
\dot{y}= q=P_{3}(q,y),
\end{equation}
and
\begin{equation}\label{46}
\dot{q}= [-f'(y)(\varepsilon_5+q^{2})]_{unperturbed}-[\phi'(y)\varepsilon_5 +\delta(f'(y)e^{2f(y)})e^{-2f(y)}(\varepsilon_5+q^{2})]_{perturbed}=Q_{3}(q,y).
\end{equation}
In the case of \emph{null geodesics}, $\varepsilon_5 = 0$. Substituting in equation (\ref{46}), we find that we get back equation (\ref{44}). So we can say that both homothetic motion as well as conformal motion produce the same effect on the null geodesics.

But the motion in the case of \emph{timelike geodesics} depends on the the factor $\phi$, and as a result the critical point may change. It is to be noted that to get a valid perturbation at every point along the extra dimension, $\phi(y)$ must be a smooth function. To obtain the critical point of the perturbed system for timelike geodesics when $f'(y)e^{2f(y)}$ is conserved (i.e for a Mashhoon-Wesson-type metric), we substitute
\begin{eqnarray}
    P_{3}(q,y) &=& 0, \nonumber \\
    \\
     Q_{3}(q,y) &=& 0. \nonumber
\end{eqnarray}
From the first equation we obtain $q = 0$, and from the second equation we find that the critical point will exist if we have a point $y = l$, such that
\begin{eqnarray}\label{47}
  -f'(l) &=& \phi'(l).
\end{eqnarray}
For instance, if we choose $\phi(y) = c_{1}y^{n},$ where $c_{1}$ is a constant and $n$ is an integer, and examine the critical point, we obtain $y=(\frac{1}{-c_{2}n})^{\frac{1}{n}} $ and $q= 0$, where $c_2$ is another constant. To determine the stability and the eigenvalues of this system, we calculate the partial derivative of $P_3$ and $Q_3$ with respect to $y$ and $q$ at the critical point. Thus we obtain
\begin{equation}\label{48}
   \alpha_{1}= (\partial P_{3} /\partial y)|_{q= 0,y=(\frac{1}{-c_{2}n})^{\frac{1}{n}}}= 0,
   \end{equation}
   \begin{equation}\label{49}
   \beta_{1}= (\partial P_{3} /\partial q)|_{q= 0,y=(\frac{1}{-c_{2}n})^{\frac{1}{n}}}= 1,
   \end{equation}
   \begin{equation}\label{50}
  \zeta_{1} = (\partial Q_{3} /\partial y)|_{q= 0,y=(\frac{1}{-c_{2}n})^{\frac{1}{n}}}= - \left(f''\left(\frac{1}{-c_{2}n}\right)^{\frac{1}{n}}+\phi''\left(\frac{1}{-c_{2}n}\right)^{\frac{1}{n}}\right),
   \end{equation}
   \begin{equation}\label{51}
   \varsigma_{1}= (\partial Q_{3} /\partial q)|_{q= 0,y=(\frac{1}{-c_{2}n})^{\frac{1}{n}}}= 0.
\end{equation}
To judge the character of the critical points, let us define
\begin{equation}\label{52}
  p =  \alpha_{1}+\varsigma_{1}= 0,
\end{equation}
\begin{equation}\label{53}
  r =\alpha_{1}\varsigma_{1} -\beta_{1}\zeta_{1} =   \left(f''\left(\frac{1}{-c_{2}n}\right)^{\frac{1}{n}}+\phi''\left(\frac{1}{-c_{2}n}\right)^{\frac{1}{n}}\right),
\end{equation}
\begin{equation}\label{54}
  \Delta = p^{2}-4r = -4\left(f''\left(\frac{1}{-c_{2}n}\right)^{\frac{1}{n}}+\phi''\left(\frac{1}{-c_{2}n}\right)^{\frac{1}{n}}\right).
\end{equation}

\subsubsection{When $n$ is negative and $c_{2}$ is positive}
Under this condition, $\phi$ becomes a decaying function, and $r$ assumes a negative value. Thus the equilibrium point is a center and the solutions near the equilibrium point acquire the topology of a circle. In that situation, the phase portrait consists of closed curves. They describe the motion of particles oscillating indefinitely about the perturbed hypersurface.

\subsubsection{When $n$ is positive and $c_{2}$ is negative}
In this case $\phi$ becomes diverging in nature, but we will still get the same type of critical point i.e a center.

\bigskip

\begin{figure}[ht]
\includegraphics[height= 1.5in]{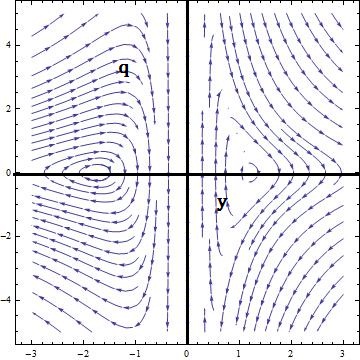}
\includegraphics[height= 1.5in]{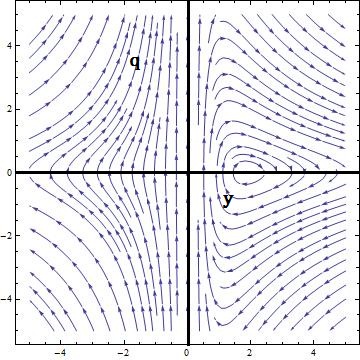}
\caption{Perturbed dynamical system for conformal motion when (i) $\phi = y^{-3}$ (ii) $\phi = y^{-4}$}
\label{Fig 3}
\end{figure}

\begin{figure}[ht]
\includegraphics[height= 1.5in]{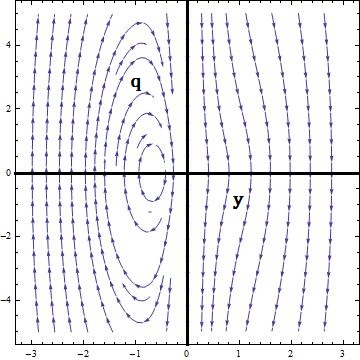}
\includegraphics[height= 1.5in]{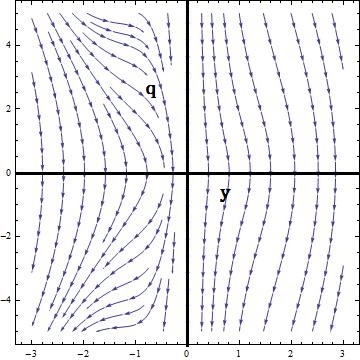}
\caption{Perturbed dynamical system for conformal motion when (i) $\phi = y^{3}$ (ii) $\phi = y^{4}$}
\label{Fig 4}
\end{figure}

\textbf{Comparison of the above two cases for timelike geodesics:} Although the critical point turns out to be a center for both the above two cases, representing stable equilibrium conditions, the position of the critical point will be different in the two cases. For the first case the equilibrium point will be far away from the brane, whereas for the second case it will be nearer to the brane as shown in Fig.~\ref{Fig 3} and Fig.~\ref{Fig 4}. We have chosen integral values for the power of $y$ in $\phi$ (since $\phi$ is a smooth function). When we choose a positive integer as in Fig.~\ref{Fig 4}, the perturbed trajectories diverge. On the contrary, when we choose a negative $n$, the perturbed trajectories die out as we move far away from the brane.


\section{Geodesic motion when $G^{t}_{y} \neq 0$}
In this case, matter fields are not confined to the brane, so the non-diagonal element of the Einstein tensor satisfies the condition
\begin{equation}\label{55}
\frac{3}{e^{2f}} \left(   \frac{\dot{T}}{T}f^{\prime} \right)\neq 0.
\end{equation}
The motion along the extra dimension can be described with the help of dynamical variables `y' and `q' defined as
\begin{equation}\label{56}
 \frac{dy}{d\lambda} = q =P_{4}(y,q),
\end{equation}
\begin{equation}\label{57}
  \frac{dq}{d\lambda}= -\frac{f'}{T^{2}}(\varepsilon_5+ T^{2}q^{2})-{2}c\frac{dt}{d\lambda}q = Q_{4}(y,q,t),
\end{equation}
where $P_{4}$ and $Q_{4}$ are well defined functions of $y$ and $q$. The most important point is that the motion along the extra dimension is not independent i.e we cannot dissociate the motion along the extra dimension from the coordinates on the brane. The motion not only depends on $t$ but also on the $\frac{dt}{d\lambda}$ term. It is very much clear that the system is nonautonomous and for a nonautonomous system, generally each point of the phase space is intersected by many distinct trajectories \cite{rs5a}. So to avoid any ambiguity we restrict our study to a particular hypersurface $y= y_{0}$.

For timelike geodesics we have $\varepsilon_{5}$ = 1. The critical point of the dynamical system can be found by solving the equations
\begin{center}
   $ P_{4}(y,q)=0 = Q_{4}(y,q)$.
\end{center}
From equation (\ref{56}) we  have $q=0$, and hence $y=$ constant at the critical point. Thus to find a critical point we have to assume that there is a point $y=l$ such that $f'(l)=0$ at that point (in view of equation (\ref{57})). These solutions, representing the equilibrium points of the dynamical system in the phase plane, correspond to curves which lie entirely in the $y= l$ hypersurface. For the particular induced geometry, the hypersurface containing the critical point is ``totally geodesic". However, a $y=\textrm{constant}$ hypersurface which does not contain the critical point, does not enjoy this special status. It is to be noted that $Z_{2}$ symmetry will hold good with respect to this hypersurface \cite{dahia2}. To find the stability of the critical point we have to calculate the following terms, and the dynamical system can be represented as a suitable combination of the these terms:
\begin{equation}\label{58}
\alpha_{2}=(\partial P_{4}/\partial y)|_{q=0,y=l} = 0,
\end{equation}
\begin{equation}\label{59}
\beta_{2}=(\partial P_{4}/\partial q)|_{q=0,y=l} = 1,
\end{equation}
\begin{equation}\label{60}
\zeta_{2}=(\partial Q_{4}/\partial y)|_{q=0,y=l} = \frac{f''(l)}{T^{2}}\varepsilon_{5} ,
\end{equation}
\begin{equation}\label{61}
\varsigma_{2}=(\partial Q_{4}/\partial q)|_{q=0,y=l} = 2c \frac{dt}{d\lambda}.
\end{equation}
\begin{figure}[ht]
\includegraphics[height=2.0in]{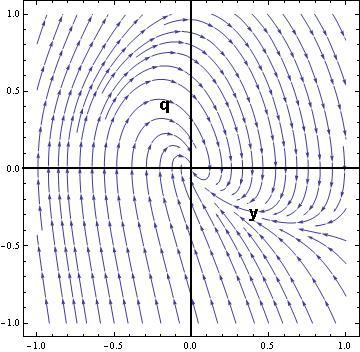}
\caption{Dynamical system for growing warp factor for $G^{t}_{y} \neq 0$}
\label{Fig 5}
\end{figure}
\begin{figure}[ht]
\includegraphics[height=2.0in]{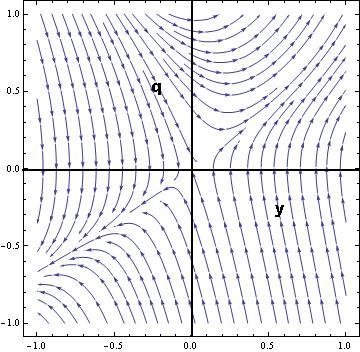}
\caption{Dynamical system for decaying warp factor for $G^{t}_{y} \neq 0$}
\label{Fig 6}
\end{figure}

We like to mention that $\frac{dt}{d\lambda}$ is always positive, as particles are traveling forwards in time, and according to our choice, the extra dimensional scale factor is of decaying type, so that $\frac{\dot{T}}{T}< 0 $, in order to ensure stabilization of the bulk after sufficient interval of time. Thus the phase portrait will be as in Fig.~\ref{Fig 5} and Fig.~\ref{Fig 6} for growing and decaying warp factors respectively. But if $y\neq l$ such that $f'(l)\neq0$, then there will not be any critical point. Such a situation is encountered in the case of the Mashhoon-Wesson-type metric. On the other hand, for null geodesics, the system has infinite number of critical points along the extra dimension.

\subsection{Perturbation as homothetic motion}

The motion along the extra dimension can again be described with the help of dynamical variables `y' and `q' as
\begin{equation}\label{62}
 \frac{dy}{d\lambda} = q = P_{5}(q,y),
\end{equation}
\begin{eqnarray*}
  \frac{dq}{d\lambda} = \left[ -(\frac{f'}{T^{2}}(\varepsilon_5+ T^{2}q^{2})+2c\frac{dt}{d\lambda}q ) \right]_{unperturbed} \qquad \qquad \qquad \qquad
\end{eqnarray*}

\begin{equation}\label{63}
 \qquad \qquad \qquad \qquad  +\left[ -\delta(f'(y)e^{2f(y)})e^{-2f(y)}(\frac{\varepsilon_5}{T^2}+q^{2}) -\delta(2c\frac{dt}{d\lambda}\frac{dy}{d\lambda})\right]_{perturbed} = Q_{5}(q,y).
\end{equation}

So to determine the critical points we have to solve the equations
\begin{center}
   $ P_{5}(y,q)=0 = Q_{5}(y,q)$.
\end{center}

The condition $P_{5}(q,y)= 0$ implies that $q = 0$. With this constraint in effect, if we try to solve $Q_{5}(0,y)= 0$, then we obtain
\begin{equation}\label{64}
   -\frac{f'}{T^{2}}(\varepsilon_5) -\delta(f'(y)e^{2f(y)})e^{-2f(y)}\frac{\varepsilon_5}{T^2}-(2c\frac{dt}{d\lambda}\delta q) = 0.
\end{equation}
In the case of null geodesics, the first two terms of equation (\ref{64}) will reduce to zero because $\varepsilon_{5} = 0$ . So to obtain a critical point, we have to assume that $\delta q = 0$ at $q = 0$, which means that $q = 0$ is the extremum of $q$. Therefore unlike the case of unperturbed motion, where we obtained an infinite number of critical points, here we do not obtain infinite number of critical points. Rather there may be a finite number of critical points depending on the number of extrema of $q$ at the point $q = 0$, for the different trajectories intersecting at $q=0$. If $q$ has no extremum at $q=0$, then the system will not have any critical point.

On the other hand, for the Mashhoon-Wesson-type metric the second term of equation (\ref{64}) becomes zero and the first term will be $-\frac{\varepsilon_{5}}{y T^{2}}$. Thus for timelike geodesics, the solution of equation (\ref{64}) can be written as
 \begin{eqnarray}
   y &=& \textrm{constant} / \delta q.
\end{eqnarray}

Thus if $q$ does not possess any extremum value at $q=0$, then the system will have no critical point, but if $q$ has an extremum along the extra dimension then we may obtain a finite number of critical points for the same reason as mentioned earlier. As $\delta q$ is very small, the position of the critical point will occur far from the brane.

\subsection{Perturbation as conformal motion}

In the case of conformal motion, we can choose $\phi = \phi(y,t)$. During a conformal motion the dynamical system can be written as
\begin{equation}\label{65}
 \frac{dy}{d\lambda} = q = P_{6}(q,y),
\end{equation}

\begin{eqnarray*}
  \frac{dq}{d\lambda} = \left[ -(\frac{f'}{T^{2}}(\varepsilon_5+ T^{2}q^{2})+2c\frac{dt}{d\lambda}q ) \right]_{unperturbed} \qquad \qquad \qquad \qquad \qquad
\end{eqnarray*}
\begin{equation}\label{66}
  \qquad \qquad \qquad +\left[-\phi'(y)\varepsilon_5 -\delta(f'(y)e^{2f(y)})e^{-2f(y)}(\frac{\varepsilon_5}{T^2}+q^{2}) -\delta(2c\frac{dt}{d\lambda}\frac{dy}{d\lambda})\right]_{perturbed} = Q_{6}(q,y).
\end{equation}
In this case, to find the critical point we have to solve the equation
\begin{center}
   $ P_{6}(y,q)=0 = Q_{6}(y,q)$.
\end{center}
Once again we find that the above condition implies that $q=0$, and therefore we obtain
\begin{equation}
   -\frac{f'}{T^{2}}(\varepsilon_5) -\phi'(y)\varepsilon_5 -\delta(f'(y)e^{2f(y)})e^{-2f(y)}(\frac{\varepsilon_5}{T^2})-2c\frac{dt}{d\lambda}\delta q = 0.
\end{equation}
To obtain a critical point for the null geodesics $(\varepsilon_5 = 0$), we assume that $q$ has an extremum at $q=0$. This is similar to the condition which we used earlier for homothetic motion. In the case of timelike geodesics $(\varepsilon_5 = 1$) for the Mashhoon-Wesson-type metric, the equation $Q_{6}(0,y)=0$ becomes
\begin{equation}\label{67}
   -\frac{f'}{T^{2}} -\phi'(y) -2c\frac{dt}{d\lambda}\delta q = 0
\end{equation}
In this case there may be specific critical points depending on the nature of the function $\phi'(y)$. Unlike the case of homothetic motion, here we need not set any constraint on $q$ to get a critical point.

\section*{CONCLUSIONS}
The results obtained in this piece of investigation are now listed below:

$\bullet$ In the first part of this paper we have studied the geodesics along the extra dimension in a 5D warped product spacetime with spacelike dynamic extra dimension. First we obtained the condition to get the critical point in the unperturbed situation when the cross-diagonal term of the Einstein tensor vanishes. Critical points reveal the nature of stability of the system and the confinement. Here we observed that the nature of the critical points for the timelike geodesics, depend on the warping function. But critical points of null geodesics are independent of the warping function and are degenerate.

$\bullet$  Subsequently we introduced the concept of an energy function along the phase path. We studied the motion along the 5th dimension during a coordinate translation occurring due to the displacement of the brane as a result of its interaction with the bulk graviton. We found that additional force terms enter the the geodesic equation due to the displacement of the brane. We also determined the change in the energy function due to such brane displacements for the case of growing and decaying warp factors.

$\bullet$ Next we considered metric perturbation and derived the perturbed geodesic equation. We have analyzed the system for two different situations, first when the perturbation generates homothetic motion and second when the perturbation produces conformal motion. In the case of homothetic motion, the critical point for a null geodesic remains unchanged. During the study of timelike geodesics, we found that the critical point turns out to be different from the unperturbed condition as the deviation of the function $f'(y)e^{2f(y)}$ becomes numerically equal to the function at the particular point in phase space. But such a condition is not admissible when the perturbation is very small. On the other hand if we assume that $f'(y)e^{2f(y)}$ is a conservative term, then we can have $\delta(f'(y)e^{2f(y)}) = 0$ . This constraint forced us to determine the warping function associated with the dynamical system under such a condition. We found that the warping function is of the type $f(y)= \frac{1}{2}\log(Cy)$. This is the warping function associated with the Mashhoon-Wesson metric. It is known that a dynamical system under Mashhoon-Wesson metric does not have any critical point in the unperturbed condition \cite{DGR}, but here in addition to that we found that it does not have any critical point even in the case of homothetic motion.

$\bullet$ For the case of perturbation in the form of conformal motion, we obtained the critical point for null geodesics. Here also the critical points are degenerate. It is clear from the above results that the nature of the critical point of a null geodesic remains the same for perturbed and unperturbed conditions for both homothetic and confornal motions. On the contrary, timelike geodesics for a Mashhoon-Wesson metric shows some important features. It has been shown that the nature of the critical point depends on the nature of $\phi$. To ensure the continuity of the effect of perturbation, we have assumed $\phi$ to be a smooth function. Further we assumed integral powers in the expression $\phi = c_{1}y^{n}$. For a positive value of $n$, $\phi$ diverges as we move away from the brane. So to curb the effect of perturbation, a negative value for the power $n$ appears to be a better choice. The number of critical points and the distance of those points from the brane depends on $n$.

$\bullet$ In the last section we examined the effect of perturbation on geodesic motion with a non-vanishing cross-diagonal component of the Einstein tensor. Here we are unable to decompose the geodesic equation along the extra dimension from the $\frac{dt}{d\lambda}$ term. Thus the critical point and its nature depends on this term. In the perturbed situation, the critical points for null geodesics may not be degenerate. Rather we may get a finite number of critical points if $q$ possesses an extremum at $q=0$. Similarly for timelike geodesics we may get critical points. But in the case of timelike geodesics, even if there are critical points, those will be far away from the brane. During conformal motion we observed that the presence of a critical point does not require any constraint on $q$, rather the nature of $\phi$ determines the position of the critical point. Another important point is the presence of the term $T^{2}$ in equation (\ref{67}). It indicates that the position of the critical point will change with time.

In this study, we have differentiated between a local and a global perturbation. In the case of a local perturbation in the form of a brane displacement due to interaction with the graviton, we have defined an energy function which can be used as a measure of the effect of perturbation. What remains to be done is the calculation of a similar parameter in the case of the global perturbation, which can be used to differentiate between the effects of the local and the global perturbations.

\section*{Acknowledgments}
SG thanks IUCAA, India for an associateship. PB gratefully acknowledges the facilities available at the Department of Physics, St. Xavier's College (Autonomous), Kolkata and the Relativity and Cosmology Research Center, Jadavpur University. We thank Prof. Subenoy Chakraborty for his suggestions.

\end{document}